\documentclass[10pt,a4paper,UKenglish]{article}
\usepackage{a4wide} 

\usepackage{authblk} 
\usepackage{cite}
\usepackage[utf8]{inputenc}
\usepackage{amssymb,amsthm,amsmath,amsfonts}
\usepackage{xspace}
\usepackage{nameref,hyperref,cleveref,thm-restate}
\hypersetup{
    colorlinks,
    citecolor=black,
    filecolor=black,
    linkcolor=blue,
    urlcolor=black
}
\usepackage{stmaryrd}
\usepackage{mathrsfs}
\usepackage{float}
\usepackage{graphicx}
\usepackage{url}
\usepackage{numprint}
\usepackage{verbatim}
\usepackage{wasysym}
\usepackage{listings}
\usepackage{multicol}

\usepackage{color}
\usepackage[svgnames]{xcolor}

\newcommand{\qc}{QuickChick}

\definecolor{redcomm}{rgb}{0.8,0,0}

\lstdefinelanguage{Coq}
{
 literate=
  {forall}{{$\forall\ $}}1
  {exists}{{$\exists\ $}}1
  {=>}{{$\Rightarrow\ $}}1
  {==>}{{$\Longrightarrow\ $}}1
  {>->}{{$\rightarrowtail\ $}}2
  {<-}{{$\leftarrow\ $}}1
  {<->}{{$\leftrightarrow\ $}}1
  {->}{{$\to\ $}}1
  {<=}{{$\leq$}}1
  {>=}{{$\geq\ $}}1
  {\/\\}{{$\wedge\ $}}1
  {|-}{{$\vdash$}}1
  {\\\/}{{$\vee\ $}}1
  {~}{{$\sim$}}1,
 basicstyle=\small\ttfamily,
 sensitive=true,
 morecomment=[l][\color{redcomm}]{//},
 morecomment=[s][\color{redcomm}]{(*}{*)},
 morestring=[b],
 stringstyle=\color{black},
 showstringspaces=false,
 firstnumber=\thelstnumber,
 numberstyle=\small,
 numberblanklines=true,
 showspaces=false,
 showtabs=false,
 xleftmargin=0pt,
 xrightmargin=0pt,
 emph=
 {[1]
  Record,Definition,Inductive,Parameter,Parameters,Axiom,Axioms,Fixpoint,Conjecture,Variable,
  Variables,Hypothesis,Hypotheses,CoInductive,Theorem,Lemma,CoFixpoint,Let,Notation,Eval,Goal,
  Qed,Defined
  },
 emphstyle={[1]\color{red}},
 emph=
 {[2]
  Module,Set,Type,Prop,fix,with,cofix,forall,fun,let,if,then,else,match,end,return,as,for,in,
  nat,list,compute,unfold,intro,intros,assert,omega,decompose,try,simpl,or,and,subst,true,false,True,False,
  firstorder,nth
  },
 emphstyle={[2]\color{blue}}
}

\lstdefinestyle{coqinline}{language={Coq},%
  basicstyle=\normalsize\ttfamily,%
}

\lstnewenvironment{codecoq}{\lstset{language=Coq,numbers=none}}{\smallskip}

\lstdefinestyle{coq}{language={Coq},%
  rangeprefix=(*-\ ,
  rangesuffix=\ *),
  numbers=none,
  captionpos=b,
  includerangemarker=false,
  columns=flexible
}
 \newcommand{\coq}[1]{\lstinline[style=coqinline]{#1}}
\newcommand{\mathcoq}[1]{\text{\coq{#1}}}

\lstdefinelanguage{prettyProlog}  %
{
 captionpos=b,
 basicstyle=\ttfamily,
 sensitive=true,
 morecomment=[l][\color{gray}]{//},
 morecomment=[s][\color{gray}]{/*}{*/},
 morestring=[b],
 stringstyle=\color{black},
 showstringspaces=false,
 firstnumber=\thelstnumber,
 numberstyle=\small,
 numberblanklines=true,
 showspaces=false,
 showtabs=false,
 xleftmargin=0pt,
 xrightmargin=-20pt,
 emph=
 {[1]
  halt,is,fail,member
  },
 emphstyle={[1]\color{red}},
 emph=
 {[2]
   line\_endo,line_endo,iterate,line,in,write_coq
  },
 emphstyle={[2]\color{blue}}
}

\lstdefinestyle{prolog}
{
 rangeprefix=/*-\ ,
 rangesuffix=\ */,
 language=Prolog,
 alsolanguage=prettyProlog,%
 basicstyle=\scriptsize\ttfamily,
}

\lstdefinestyle{prologinline}{
  language={Prolog},%
  alsolanguage=prettyProlog,%
  basicstyle=\ttfamily,%
}

\newcommand{\coqlisting}[2]{
\lstinputlisting[style=coq,
linerange={BEGIN#1-END#1},
includerangemarker=false]{#2}}

\theoremstyle{plain}
\newtheorem{proposition}{Proposition}

\begin{document}


\title{Pragmatic isomorphism proofs between Coq representations: application to lambda-term families}


\author[1]{Catherine Dubois}
\author[2]{Nicolas Magaud}
\author[3]{Alain Giorgetti}

\affil[1]{Samovar, ENSIIE, 1 square de la r\'esistance, 91025 Evry, France}
\affil[2]{Lab. ICube UMR 7357 CNRS Universit\'e de Strasbourg, 67412 Illkirch, France}
\affil[3]{FEMTO-ST Institute, Univ. Bourgogne Franche-Comt\'e, 25030 Besan\c{c}on, France}

\date{}






\maketitle

\begin{abstract}
There are several ways to formally represent families of data, such as lambda
terms, in a type theory such as the dependent type theory of Coq.
Mathematical representations are very compact ones and usually rely on the use
of dependent types, but they tend to be difficult to handle in practice. On the
contrary, implementations based on a larger (and simpler) data structure
combined with a restriction property are much easier to deal with.

In this work, we study several families related to lambda terms, among which
Motzkin trees, seen as lambda term skeletons, closable Motzkin trees,
corresponding to closed lambda terms, and a parameterized family of open lambda
terms. For each of these families, we define two different representations, show
that they are isomorphic and provide tools to switch from one representation to
another. All these datatypes and their associated transformations are
implemented in the Coq proof assistant. Furthermore we implement random
generators for each representation, using the QuickChick plugin.
\end{abstract}


\section{Introduction}

Choosing the most appropriate implementation of mathematical
objects to perform computations and proofs is
challenging. Indeed, efficient (well-suited for computations)
representations are often difficult to handle when it comes to proving
properties of these objects.  Conversely, well-suited representations
for proofs often have fairly poor performances when it comes to
computing.  The simplest example is the implementation of natural
numbers. Using a unary representation, proofs  (especially inductive reasoning)
are easy to carry out but computing is highly inefficient.  Using a
binary representation makes computations faster, however it is more
difficult to use reasoning principles such as the induction principle on
natural numbers.  

In the field of $\lambda$-calculus, representations which
are the closest to mathematics are usually implemented using dependent types. This
makes them easily readable and understandable by
mathematicians. However it is rather challenging and requires a strong
background in functional programming and theorem proving to handle them smoothly.  Representations based on a larger type (a
non-dependent one) and a restriction property are easier to handle
in practice, but less intuitive.  In addition, one needs to take extra
care to make sure that the combination of the larger type and the
restriction property exactly represents the expected objects.

Overall there is no perfect representation for a given
mathematical object.
To overcome this challenge, we propose to deal
with several different isomorphic representations of families
of mathematical objects simultaneously.   To do that, we present a rigorous
methodology to partially automate the construction of the transformation
functions between two isomorphic representations and prove these transformations correct. We apply these
techniques to some families of objects related to $\lambda$-calculus, namely
closable Motzkin trees, uniquely closable Motzkin trees and $m$-open $\lambda$-terms.

Our first examples revisit an article of Bodini and Tarau~\cite{lopstr17} in
which they define Prolog generators for closed lambda terms and their skeletons
seen as Motzkin trees, efficient generators for closable and uniquely closable
skeletons and study their statistical properties.
Our contributions are to formalize in Coq these different notions, prove the
equivalence of several definitions that underlie the generators designed by
Bodini and Tarau, and write random generators to be used with
QuickChick~\cite{Pierce:SF4}. We then extend the discourse to a parameterized
family of open $\lambda$-terms, named $m$-open $\lambda$-terms. All
the considered representations, transformations between isomorphic
representations and isomorphism proofs are formalized\footnote{The Coq code, for
Coq~8.10.2 and QuickChick~1.2.1, is freely distributed in CUT 2.5, downloadable
at \url{http://members.femto-st.fr/alain-giorgetti/en/coq-unit-testing}. It also
works up to Coq~8.16.}  in the Coq proof assistant~\cite{BC04,coqmanual}. We
propose some generic tools to help setting up the correspondence between two
isomorphic types more easily.  We hope such a methodology could be reused to
deal with other families of objects, having different and isomorphic
representations.

\subparagraph{Related Work.}
Dealing with various isomorphic representations of the same mathematical objects
is a common issue in computer science. Research results span from
theoretical high-level approaches such as homotopy type theory \cite{hottbook}
or
cubical type theory \cite{DBLP:journals/flap/CohenCHM17} to more
pragmatic proposals such as ours.  
In the context of formal specifications
and proofs about mathematical concepts, several frameworks have been proposed to deal with
several types and their transformation functions. A seminal work on changing
(isomorphic) data representation was implemented by
Magaud~\cite{DBLP:conf/tphol/Magaud03}  as a plugin for Coq in the early 2000s.
In this approach, the transformation functions were provided by the user and
only the proofs were ported. Here, we aim at helping the programmer to write the
transformation functions as well as their proofs of correctness.
In~\cite{DBLP:conf/cpp/CohenDM13}, Cohen et al. focus on refining from abstract
representations, well-suited for reasoning, to computationally well-behaved
representations. In our work, both representations are considered of
equal importance, and none of them is preferred.
Finally, our work is closely related to the concept of views, introduced by
Wadler in~\cite{DBLP:conf/popl/Wadler87} and heavily used in the
dependently-typed programming language
Epigram~\cite{DBLP:journals/jfp/McBrideM04}.  In this approach, operations are
made independent of the actual implementation of the types they work on.
Pattern-matching on an element of type $A$ can be carried out following the
structure of the type $B$ provided $A$ and $B$ are isomorphic types.
 The correspondence functions we shall implement in this article provide an
example of a concrete implementation of views.

Regarding random generators and enumerators, Paraskevopoulou et al.~\cite{10.1145/3519939.3523707}
recently proposed a new framework,  on top of the QuickChick
testing tool for Coq. It allows to automatically derive such
generators by extracting the computational contents from inductive relations.  

\subparagraph{Paper Outline.} In Sect.~\ref{sec:methodology}, we present a general
methodology and interfaces to capture all the features of two representations of
a given family of objects, and to switch easily from one representation to the
other.  In Sect.~\ref{sec:closable}, we show how our approach applies
to representations of closable Motzkin trees -- that are the skeletons
of closed $\lambda$-terms -- and to representations of 
uniquely closable Motzkin trees. In Sect.~\ref{sec:mopen}, we adapt our
approach to the parameterized family of $m$-open
$\lambda$-terms. In Sect.~\ref{bt17-sec}, several applications of the presented isomorphic types are
exposed. In Sect.~\ref{sec:conclusion}, we draw some
conclusions and present some promising perspectives.

\section{Specifying Families using Two Different Representations}
\label{sec:methodology}

As we shall see with examples related to pure $\lambda$-terms, a family of
mathematical objects can usually be defined formally in two different but
equivalent ways: either using an inductive datatype, possibly dependent, or
using a larger non dependent datatype, together with a restriction property. In
this section, we summarize which elements are required to specify the two
datatypes and their basic properties. We then show how to derive the isomorphism
properties automatically.  One of these isomorphism properties can
always be
derived automatically, using a generic approach based on a functor, whereas the other one, which
relies on a proof by induction on the data, is carried out using
Ltac. Although the Ltac code aims at being as generic as possible, the
only warranty we can provide is that it works for all of our examples. 

\vspace{-8pt}

\subsection{Types}

A \emph{restricted type} (\coq{T},\coq{is_P}) is a dependent pair defined by a type
\coq{T : Type}, called its \emph{base type}, and a predicate \coq{is_P : T ->
Prop}, called its \emph{restriction} or \emph{filter}.
The restricted type (\coq{T},\coq{is_P}) is intended to represent the
inhabitants of \coq{T} satisfying the restriction \coq{is_P}. For practical
reasons, these two objects are encapsulated together as a record type 
\coq{rec_P}
 isomorphic to the $\Sigma$-type $\{x:\mathcoq{T}~ |~ \mathcoq{is\_P}~ x\}$.%
\begin{lstlisting}[style=coq] 
Record rec_P := Build_rec_P {
  P_struct :> T;
  P_prop : is_P P_struct
}.
\end{lstlisting}

In addition to this practical type, we 
assume that we also have another (possibly dependent type) \coq{P} for the same family of
objects. This type is usually closer to the way mathematicians would define such
objects. However, it may be less convenient to handle in practice (e.g. when proving
in a proof assistant such as Coq) and thus we shall prefer using the larger type
\coq{T} and the restriction \coq{is_P} rather than the type \coq{P}
when programming operations and proving lemmas on such a family.

\vspace{-8pt}
\subsection{Transformations and their properties}

Once the datatypes \coq{T} and \coq{P} and the filter \coq{is_P} are defined, we
build the expected isomorphisms as two transformation functions \coq{rec_P2P}
(from $\mathcoq{rec\_P} \equiv \{x:\mathcoq{T}~ |~ \mathcoq{is\_P}~ x\}$ to $P$)
and \coq{P2rec_P} (from $P$ to $\mathcoq{rec\_P} \equiv \{x:\mathcoq{T}~ |~
\mathcoq{is\_P}~ x\}$). The first function \coq{rec\_P2P} can be defined as
follows, with an auxiliary function \coq{T2P : forall (x:T), is_P x -> P}
transforming any element $x:$ \coq{T} that satisfies $\mathcoq{is\_P}$ into an
element of \coq{P}.
\begin{lstlisting}[style=coq]
Definition rec_P2P m := T2P (P_struct m) (P_prop m).
\end{lstlisting}
To define the reverse transformation \coq{P2rec\_P}, we first
implement a function \coq{P2T} from \coq{P} to \coq{T} and then prove that the image of
any $x$ by \coq{P2T} satisfies the predicate \coq{is_P}, i.e. we prove the following lemma:
\begin{lstlisting}[style=coq]
Lemma is_P_lemma: forall v, is_P (P2T v).
\end{lstlisting}
\noindent Then the
transformation  \coq{P2rec\_P} can be defined as follows:
\begin{lstlisting}[style=coq]
Definition P2rec_P (x:P) : rec_P := Build_rec_P (P2T x) (is_P_lemma x).
\end{lstlisting}

\vspace{-8pt}
\subsection{Partial automation of specification and proofs}

In order to automate some parts of the process, we provide an
abstract definition of the minimum requirements for the two involved
types, as shown in the module type declaration \coq{family} 
reproduced in the following code snippet.

\lstinputlisting[style=coq,
linerange={BEGINfamily-ENDfamily},
includerangemarker=false]{familyInterface.v}

We assume that we have the type \coq{T} and a restricting predicate \coq{is\_P}
as well as the type \coq{P}.  We also provide two conversion functions \coq{T2P}
and \coq{P2T}, together with two proofs: a proof \coq{is\_P\_lemma} that
\coq{is\_P} holds for all images \coq{(P2T v)} of the inhabitants \coq{v} of
\coq{T}, and a proof \coq{P2T_is_P} that for all inhabitants \coq{t : T}
satisfying the predicate \coq{is\_P}, \coq{P2T} is a left inverse of \coq{T2P}.

Then, the roundtrip lemma \coq{P2rec\_PK} stating that \coq{P2rec\_P} is a
left inverse for \coq{rec\_P2P} can be proved automatically using the functor
\coq{equiv\_family}, reproduced in the following code snippet.
\lstinputlisting[style=coq, linerange={BEGINfunctor-ENDfunctor},
includerangemarker=false]{familyInterface.v}
The proof of \coq{P2rec_PK} is generic and only relies on the components
of the module \coq{f} which has type \coq{family}.

The proof of the other roundtrip lemma \coq{rec\_P2PK} cannot be derived abstractly using
a functor. Indeed, the argument of this lemma is an element \coq{m} of the
inductively-defined type \coq{P}. Therefore no proof can be carried out before
we have an explicit definition of \coq{P}. Once this definition is provided, the
proof of the second lemma is rather straightforward and we can be automated
using some Ltac constructs. Although the Ltac proof is not generic, it works
easily for all examples provided in this paper.  We believe that this could be
generalized to arbitrary datatypes by using some meta-programming tools such as
Coq-elpi~\cite{tassi:hal-01637063} or
MetaCoq~\cite{DBLP:journals/jar/SozeauABCFKMTW20}.

In the next subsection, we shall extend our interface and build a new
functor to automatically
generate some random generators for the two representations \coq{P}
and $\{x:\mathcoq{T}~ |~ \mathcoq{is\_P}~ x\}$ at stake.  

\vspace{-8pt}
\subsection{Random generators}

Property based testing (PBT) has become famous in the community of functional
languages. Mainly popularized by QuickCheck~\cite{quick2000} in Haskell, PBT is
also available in proof assistants. In Coq, the  random testing plugin
QuickChick~\cite{Pierce:SF4} allows us to check the validity of executable
conjectures with random inputs, before trying to write formal proofs
of these conjectures.  QuickChick is mainly a generic framework
providing combinators to write testing code, in particular random generators,
and also to prove their correctness.

Our general framework also provides guidelines to develop random generators for
all the datatypes under study.
Generators, either user-defined or automatically derived by \qc{}, have a type \coq{G}~\coq{Ty}
where \coq{Ty} is the type of the generated values and \coq{G} is an instance of
the Coq \coq{Monad} typeclass. They are usually parameterized by a natural
number \coq{n} which controls their termination (called \textit{fuel} in
the Coq community). It may also serve as a bound on the depth of the generated
values, even if it is not always guaranteed.

Let us assume that a random generator of values of type \coq{T}, named
\coq{gen\_T} : \coq{nat -> G\ T}, is available. In our context, we are mainly interested in the following generators: (i) a generator of values of type \coq{T} satisfying the property \coq{is_P}, (ii) a generator of values of type \coq{rec_P} satisfying the property \coq{is_P}, (iii) a generator of values of type \coq{P}. 
Thanks to QuickChick and the bijections we have previously defined, they can be obtained quite easily, using three new functors explained below.
All these generators come in a sized version, i.e. they are parameterized with a natural
number which is randomly chosen, when used with a  \qc{} test command.

The first functor we propose, \coq{generators_family1}, allows the definition of the random
generator \coq{gen\_filter\_P} which implements the strategy
\textit{generate and test}. It can be obtained when are available an executable version of the predicate \coq{is_P}, named \coq{is_Pb},  and a proof of decidability of  \coq{is_P}, named \coq{is_P_dec}. A value  \coq{default_P} of the considered family - which is guaranteed by a proof \coq{default_is_P} - is also required.  

\lstinputlisting[style=coq,
linerange={BEGINgen1-ENDgen1},
includerangemarker=false]{familyInterface.v}

The random
generator \coq{gen\_filter\_P} randomly produces a value \coq{val} of type \coq{T} thanks to
\coq{gen\_P} and checks whether \coq{is\_Pb}~\coq{val} is true, in which case it
outputs \coq{val}. Otherwise, it discards the value and tries again. If the
maximum number of tries \coq{filter_max} %
is reached, it yields the provided default value \coq{default\_P}.

The two next functors can be used to derive a random generator for one family representation from that of the alternative representation. When the random generator \coq{gen\_P} of values of type \coq{P} is available, using the functor \coq{generators_family3} shown below, we can obtain a random generator of values of  type \coq{rec_P}, i.e. a value of type \coq{T} and a proof that it satisfies the property \coq{is_P} (thanks to the functions and lemmas derived using \coq{equiv_family}). The functor \coq{generators_family2} (omitted here) does the opposite job. 
\lstinputlisting[style=coq,
linerange={BEGINgen3-ENDgen3},
includerangemarker=false]{familyinterface.v}

In the next section, we shall see how to instantiate our framework
with two different representations of closable Motzkin trees  and
uniquely closable Motzkin trees, to automatically prove the equivalence between the
representations and to automatically derive random generators.  

\section{Two Instances: Closable Motzkin Trees and Uniquely Closable
  Motzkin Trees}%
\label{sec:closable}

This section presents two simple examples of infinite families of objects with two
representations in Coq. These examples are presented as applications
of our formal framework, including formal proofs of isomorphisms between
representations and the design of their corresponding random generators.
Whereas our methodology applies to any pair of isomorphic datatypes, we have
chosen to focus our applications primarily on data families related to the
$\lambda$-terms from the pure (i.e., untyped) $\lambda$-calculus.

Let us briefly recall that the
$\lambda$-calculus is a universal formalism to represent computations with
functions. A \emph{(pure) $\lambda$-term} is either a variable ($x$, $y$,
\ldots), an \emph{abstraction} $\lambda x. t$, that \emph{binds} the variable
$x$ in the $\lambda$-term $t$, or a term of the form $t\, u$ for two
$\lambda$-terms $t$ and $u$. The term $\lambda x. t$ represents a function of
the variable $x$. The term $t\, u$ represents an \emph{application} of the
function (represented by) $t$ to the function (represented by) $u$. A variable
$x$ in \emph{free} in the term $t$ if it is not bound in $t$ (by some $\lambda
x$).
A \emph{closed} term is a term without free variables. Terms are considered up
to renaming of their bound variables.

The two examples come from a study for the efficient enumeration of closed
$\lambda$-terms, by Bodini and Tarau~\cite{lopstr17}, that starts from
binary-unary trees, a.k.a. Motzkin trees, that can be seen as skeletons of
$\lambda$-terms. For self-containment, all the definitions and
properties of this study that are formalized here are kindly reminded to the
reader.

A \emph{Motzkin tree} is a rooted ordered tree built from binary, unary
 and leaf nodes. Thus the set of Motzkin trees can be seen as the free
algebra generated by the constructors \coq{v}, \coq{l} and \coq{a} of respective
arity 0, 1 and 2. Their type in Coq, named \coq{motzkin}, is the following
inductive type.

\lstinputlisting[style=coq,
linerange={BEGINmotzkin-ENDmotzkin},
includerangemarker=false]{closable.v}

\vspace{-8pt}
\subsection{Closable Motzkin trees}

The \emph{skeleton} of the $\lambda$-term $t$ is the Motzkin tree obtained by
erasing all the occurrences of the variables in $t$. A Motzkin tree is
\emph{closable} if it is the skeleton of at least one closed $\lambda$-term.
As in~\cite{lopstr17}, we define a predicate for characterizing closable
Motzkin trees:

\lstinputlisting[style=coq,
linerange={BEGINis_closable-ENDis_closable},
includerangemarker=false]{closable.v}

\noindent This predicate only requires the presence of at least one occurrence
of the unary node on each rooted path of the Motzkin tree. For instance, the
tree \coq{l (a v (l v))} is closable (it is the skeleton of the closed 
$\lambda$-term $\lambda x. x (\lambda y. y)$), whereas the tree \coq{a (l v) v}
is not closable.

\begin{table}
\begin{center}
  \begin{tabular}{|l|c|c|}
    \hline
    Abstraction & Closable Skeletons & Uniquely Closable Skeletons\\
    \hline
    \coq{T} & \coq{motzkin} & \coq{motzkin} \\
    \hline
    \coq{is_P} & \coq{is_closable} & \coq{is_ucs}\\
    \hline
    \coq{P} & \coq{closable}  & \coq{ucs}\\
    \hline
    \coq{T2P} & \coq{motzkin2closable} & \coq{motzkin2ucs}\\
    \hline
    \coq{P2T}& \coq{closable2motzkin} & \coq{ucs2motzkin}\\
    \hline
    \coq{is_P_lemma} & \multicolumn{2}{c|}{automatically proved  using Ltac} \\
    \hline
    \coq{P2T_is_P}& \multicolumn{2}{c|}{automatically proved  using Ltac}\\
    \hline
    \coq{proof_irr}&\coq{proof_irr_is_closable}&\coq{proof_irr_is_ucs} \\
    \hline
    \hline
    \coq{rec_P} & \multicolumn{2}{c|}{automatically derived in the functor}\\
    \hline

    \coq{rec_P2P} & \multicolumn{2}{c|}{automatically
                    derived in the functor} \\
    \hline
 \coq{P2rec_P} & \multicolumn{2}{c|}{automatically derived in the functor}\\

    \hline
    \coq{P2rec_PK} & \multicolumn{2}{c|}{automatically derived in the functor}  \\
    \hline
    \hline
    \coq{rec_P2PK} & \multicolumn{2}{c|}{automatically proved using Ltac}\\
    \hline
  \end{tabular}
\end{center}
  \caption{\label{XXX}Two instances of the \coq{Module Type family} and the functor \coq{equiv_family} representing closable 
    Motzkin trees and uniquely closable Motzkin trees. Statements
    required in the functor \coq{Module Type family} (upper part of the
  array) are proven automatically. The roundtrip statement
  \coq{rec_P2PK} (last line of the array), which corresponds to \coq{rec_closable2closableK}
  and \coq{rec_ucs2ucsK} does not belong to the functor but can be
  proven automatically in both settings}
\end{table}
Bodini and Tarau proposed a grammar generating closable Motzkin
trees~\cite[Section 3]{lopstr17}, that we adapt in Coq as an inductive
type, named \coq{closable}.

\lstinputlisting[style=coq,
linerange={BEGINclosable-ENDclosable},
includerangemarker=false]{closable.v}
For example, \coq{La (a v (l v))} is the \coq{closable} term corresponding to the tree \coq{l (a v (l v))}.

To prove that there is a
bijection between closable Motzkin trees specified using the type
\coq{rec_closable}  and inductive objects whose type is
\coq{closable}, using our approach, we simply need to provide two
functions \coq{motzkin2closable} and \coq{closable2motzkin}.

\lstinputlisting[style=coq,
linerange={BEGINmotzkin2closable-ENDmotzkin2closable},
includerangemarker=false]{closable.v}

\lstinputlisting[style=coq,
linerange={BEGINclosable2motzkin-ENDclosable2motzkin},
includerangemarker=false]{closable.v}

Because it involves dependent pattern matching, defining directly \coq{motzkin2closable} as a function is not
immediate. However it is easily carried out interactively as a lemma, in a proof-like
manner, using the tactic \coq{fix}.

The transformation functions and the isomorphism properties between
the two types \coq{closable} and \coq{rec_closable} can then be
automatically generated, as summarized in the second column of Table~\ref{XXX}.

\subsubsection{Random Generators}
Random generators for \coq{closable} and \coq{rec_closable} have been used to test the different lemmas before proving them,
 for example the roundtrip lemma \coq{rec_closable2closableK}, which is an instance of the pattern \coq{rec_P2PK}. 
Corresponding \qc{} commands can be found in our formal development.

The generator for Motzkin trees, \coq{gen\_motzkin}, required by any of the other generators, is obtained automatically, thanks to \qc: 
\lstinputlisting[style=coq,
linerange={BEGINgen_motzkin-ENDgen_motzkin},
includerangemarker=false]{qc_closable.v}

In the context of closable Motzkin trees, the \coq{gen\_closable} generator associated to the tailored simple inductive
type \coq{closable} can be easily obtained  
using \qc{}. Thanks to the functor \coq{generators_family3}, we can derive the random generator of values of the corresponding restricted type, as it is illustrated by the following snippet of code, where \coq{closable} is an instance  of the \coq{family}, 
and \coq{fact_cl} is defined as the module  \coq{equiv_family (closable)}.

\lstinputlisting[style=coq,
linerange={BEGINgen_rec_closable-ENDgen_rec_closable},
includerangemarker=false]{qc_closable.v}

To test the \coq{motzkin2closable} function (\coq{T2P}  in the \coq{family} interface), we need a generator that produces closable Motzkin trees. It is not relevant to use the previously defined generator which we have derived from that of \coq{closable} values and thus obtained using, as a main ingredient, the function under test itself.  For that purpose,  the generator \coq{gen_filter_P} obtained by applying the functor \coq{generators_family1} can be useful, however such a generator usually discards many values to produce the required ones. A handmade generator, as \coq{gen\_closable\_struct}  defined below, is usually preferred.

As a representative of this kind of custom
generators, we expose its code in the following code snippet and explain it.
\lstinputlisting[style=coq,
linerange={BEGINgen_closable_struct-ENDgen_closable_struct},
includerangemarker=false]{qc_closable.v}
\noindent We first define an intermediate function that uses the additional parameter
\coq{k} denoting the number of \coq{l} constructors at hand. So, if both \coq{k}
and \coq{n} are equal to 0, the generator emits the default value (here \coq{l v}, stored in
\coq{default_closable}). If \coq{n} is 0
but at least one \coq{l} is available, then the generator produces the leaf
\coq{v}. When \coq{n} is not 0, again we have two treatments depending
on whether we have
already introduced the constructor \coq{l} or not. In both cases, the generator
picks  one of the several ways to produce a value -- thanks to \coq{oneOf}, and
thus either stops with a value (resp. \coq{l v} or \coq{v}), recursively
produces a closable Motzkin tree which is used to build a resulting unary
Motzkin tree, or recursively generates two closable Motzkin trees  used to
produce a binary Motzkin tree.  The final custom generator is obtained using the
previous intermediate function with \coq{k} equal to 0.

We recommand to test that this generator does produce Motzkin trees which are closable, as follows:
\lstinputlisting[style=coq,
linerange={BEGINgen_closable_struct_test-ENDgen_closable_struct_test},
includerangemarker=false]{val_qc_closable.v}

To define the proof-carrying version of the custom generator, we follow a similar scheme
but also produce a proof that the produced value \coq{mt} is closable, i.e. 
a term of type \coq{is_closable mt}. We use the \coq{Program} facility which allows us
to produce certified programs and generates proof obligations. 
Here these proof obligations are automatically solved.

\vspace{-8pt}

\subsection{Uniquely closable Motzkin trees}
\label{sec:uniquely}

A Motzkin tree is \emph{uniquely closable} if there exists exactly one closed
$\lambda$-term having it as its skeleton. 

We first define a predicate \coq{is_ucs} for characterizing uniquely closable
skeletons. This predicate specifies that a Motzkin tree is uniquely closable if
and only if there is exactly one unary node on each rooted path.

\lstinputlisting[style=coq, linerange={BEGINis_ucs-ENDis_ucs},
includerangemarker=false]{ucs.v}

This Coq predicate corresponds to the second Prolog predicate
\texttt{uniquely\-Closable2} introduced by Bodini and Tarau~\cite[Section
4]{lopstr17}, after a first Prolog predicate \texttt{uniquelyClosable1} using a
natural number to count the number of $\lambda$ binders above each leaf, instead
of a Boolean flag as here. A Coq formalization of this other characterization of
uniquely closable Motzkin trees, and a formal proof of their equivalence, are
presented in Section~\ref{prop4sec}.

We then define an inductive type \coq{ucs} that also represents uniquely
closable Motzkin trees.

\lstinputlisting[style=coq,
linerange={BEGINucs-ENDucs},
includerangemarker=false]{ucs.v}

Through we use the abbreviations \coq{ca} for \texttt{ClosedAbove} and \coq{ucs} for
\texttt{Uniquely\-Closable}, these types exactly correspond to Haskell datatypes
given in~\cite{lopstr17}.
For instance, the Motzkin tree \coq{l (a v v)} and the corresponding \coq{ucs}
term \coq{L (B V V)} represent uniquely closable skeletons. The closable tree
\coq{l (a (l v) v)} is not uniquely closable, because it is the skeleton of two
closed $\lambda$-terms, namely $\lambda x. (\lambda y. y) x$ and
$\lambda x.(\lambda y. x) x$.

Using the same infrastructure as for closable Motzkin trees, the
transformation functions and the isomorphism properties between 
the two types \coq{ucs} and \coq{rec_ucs} can be
automatically generated, as summarized in the last column of Table~\ref{XXX}.

We proceed in the same way for random generators.  Using \qc, the generator  \coq{gen\_ucs} is  
 automatically derived from the definition of the inductive types \coq{ca} and \coq{ucs}. 
The user-defined generator \coq{gen\_ucs\_struct} is very close to 
\coq{gen\_closable\_struct}. Similarly we use \coq{Program} to define the one producing values and proofs.

\section{Pure Open $\lambda$-Terms in de Bruijn Form}
\label{sec:mopen}

Let us now address the questions of formal representations and random generation
of pure open $\lambda$-terms modulo variable renaming. The definitions in
this section are not present in Bodini and Tarau's work~\cite{lopstr17}.

To get rid of variable names, we adopt de Bruijn's proposal to replace each
variable in a $\lambda$-term by a natural number, called its \emph{de Bruijn
index}~\cite{de_bruijn_lambda_1972}. When a de Bruijn index is not too high, it
encodes a variable bound by the number of $\lambda$'s between its location and
the $\lambda$ that binds it. Otherwise, it encodes a free variable.  We consider
de Bruijn indices from 0, to ease their formalization with the Coq type
\coq{nat} for natural numbers. For instance, the term $\lambda.(1~\lambda.1)$ in
de Bruijn form represents the term $\lambda x. (y~\lambda z. x)$ with the free
variable $y$.
\vspace{-8pt}
\subsection{Types}%

Consequently, open $\lambda$-terms in de Bruijn form can be represented by
unary-binary trees whose leaves are labeled by a natural number. They are the
inhabitants of the following inductive Coq type \coq{lmt} (acronym for \texttt{l}abeled
\texttt{M}otzkin \texttt{t}ree).

\lstinputlisting[style=coq,
linerange={BEGINlmt-ENDlmt},
includerangemarker=false]{open.v}

However the property of being closed cannot be defined by induction on this
definition of $\lambda$-terms. Indeed, if the term $\lambda\,t$ is closed, then
the term $t$ is not necessarily closed, it can also have a free variable. The
more general property of $m$-openness overcomes this limitation: for any natural
number $m$, the $\lambda$-term $t$ is said to be $m$\emph{-open} if the term
$\lambda\ldots\lambda\,t$ with $m$ abstractions before $t$ is closed. Whereas
the ``$m$-open'' terminology is recent~\cite{BBD19}, the notion has been studied
since 2013, by Grygiel and Lescanne~\cite{Les13,GL13}.

With the following definition, \coq{(is_open m t)} holds iff the labeled Motzkin
tree \coq{t} encodes an \coq{m}-open $\lambda$-term. This function call indeed
visits the tree \coq{t} and counts (from \coq{m}) the number of $\lambda$s
(constructor \coq{lam}) traversed so far. At each leaf (constructor \coq{var})
it checks that its de Bruijn indice \coq{i} is lower than this number \coq{m} of
traversed abstractions.%

\coqlisting{is_open}{open.v}

\noindent For instance, the tree \coq{lam
(app (var 0) (lam (var 1)))} is $0$-open (its skeleton is the closable term
\coq{l (a v (l v))}), whereas the tree \coq{lam (app (var 1) (lam (var 1)))} is
$1$-open, but not $0$-open.

Because of the extra parameter $m$, the
formal framework presented in Sect.~\ref{sec:methodology} must be adapted and
we propose a new module type \coq{param_family} together with a
functor \coq{equiv_param_family} to automatically prove one of the
roundtrip lemmas. The other one can be easily proved correct using the
same sequences of Ltac constructs as for the non dependent case.  

The following record type parameterized by \coq{m} is such that \coq{(rec_open m)}
describes \coq{m}-open terms. As previously, the first field stores the datum, 
here a labeled Motzkin tree (i.e., \coq{T} is \coq{lmt}), and the second field stores a proof that it
is \coq{m}-open.

\coqlisting{rec_open}{open.v}

It is however more natural to describe \coq{m}-open terms with a dependent
type \coq{(open m)} enclosing the condition \coq{i < m} at leaves, as follows.

\coqlisting{open}{open.v}

\vspace{-8pt}
\subsection{Transformations and their properties}

In order to switch from one representation to the other whenever
needed, we provide two functions \coq{rec_open2open m} and
\coq{open2rec_open m}, and Coq proofs for two roundtrip lemmas justifying that they are mutual
inverses.

\subparagraph{From the record type to the dependent type.}

The function \coq{rec_open2open m} from the record type (\coq{rec_open m}) to
the dependent type (\coq{open m}) is defined by
\coqlisting{rec_open2open}{open.v}
\noindent where \coq{lmt2open} is the following dependent recursive function.
\coqlisting{lmt2open}{open.v}

\noindent It is rather difficult to define this function directly. We choose to
develop it as a proof, as advocated by McBride~\cite{DBLP:conf/types/McBride00},
in an interactive manner, letting Coq handle the type dependencies. Once the
term is built, we simply revert the proof and declare it directly as a fixpoint
construction to make it look like a function, more readable and understandable
for humans than a proof script.

\subparagraph{From the dependent type to the record type.}

The process to define the inverse function \coq{open2rec_open m} from the
dependent type (\coq{open m}) to the record type (\coq{rec_open m}) is rather
different, and can be decomposed as follows. First of all, a function
(\coq{open2lmt m}) turns each dependent term \coq{t} of
 type \coq{open}~\coq{m} into a
labeled Motzkin tree. \coqlisting{open2lmt}{open.v} Then we prove automatically,
using the same Ltac constructs as for the previous examples, the following 
lemma 
 that states that the function \coq{open2lmt m}
always outputs an \coq{m}-open term.
\coqlisting{is_open_lemma}{open.v}
Once this lemma is proved, we can derive automatically the transformation 
\coq{open2rec\_open}, by using the functor \coq{equiv_param_family}.
\coqlisting{open2rec_open}{open.v}

As we did in the previous sections, we then need to prove a lemma
\coq{open2lmt_is_open} which relates the
functions \coq{open2lmt} and \coq{lmt2open}, without taking
into account the restriction property.
\coqlisting{open2lmt_is_open}{open.v}
Both lemmas are part of the interface \coq{param_family} for a parametric family, extending the interface \coq{family}. Thus,
applying the appropriate functor, we automatically derive a proof
of the first roundtrip lemma: 
\coqlisting{open2rec_openK}{open.v}

\noindent The proof of the second roundtrip lemma proceeds by induction on
\coq{x} of type \coq{open}~\coq{m}. It is
immediately proven using the Ltac constructs proposed in the previous
sections.

\coqlisting{rec_open2openK}{open.v}

\vspace{-8pt}
\subsection{Random generators}

The required generator \coq{gen\_lmt} is automatically derived by \qc{}
from the definition of the inductive type \coq{lmt}. The custom generators for
$\lambda$-terms satisfying the \coq{open m} property, with or without proofs, are
written following the same canvas as before. The generator corresponding to the
inductive type \coq{open} is no longer derived automatically by \qc, in
particular because proofs have to be inserted when using the \coq{open\_var}
constructor. However it is easy to define it manually.

\vspace{-8pt}
\subsection{Characterization of open $\lambda$-terms from their skeleton}
\label{mopen-skeleton-sec}

This subsection presents definitions and formal proofs relating Bodini and Tarau's
skeletons for $\lambda$-terms (Section~\ref{sec:closable}) with $m$-open
$\lambda$-terms introduced in this section, not present in Bodini and Tarau's
work.

The skeleton of a $\lambda$-term is the Motzkin tree obtained by erasing the
 labels at its leaves.
\coqlisting{skeleton}{BT17.v}

\noindent This function (specified by \texttt{toMotSkel} in~\cite{lopstr17})
connects Motzkin trees without labels (Sect.~\ref{sec:closable}) and Motzkin
trees with labels defined in this section.%

As the \coq{skeleton} function cannot be inverted functionality, we define a pseudo-reverse, from Motzkin
trees without labels to labeled Motzkin trees, as the following family of inductive relations (\coq{label m}), for all natural numbers
\coq{m}.

\coqlisting{label}{label.v}

The label-removing function \coq{skeleton} and the label-adding relation
\coq{label} can be used together as follows, to define a second characterization
of \coq{m}-open $\lambda$ terms among labeled Motzkin trees \coq{t}.

\coqlisting{skeleton_open}{label.v}

\noindent The proof of the following equivalence with the first characterization
(\coq{is_open}, introduced in Section~\ref{sec:mopen}) is straightforward.

\coqlisting{skeleton_is_open_eq}{label.v}

An $m_1$-open $\lambda$-term is also an $m_2$-open $\lambda$-term for all $m_2 \geq m_1$.
\coqlisting{label_mon}{label.v}

\noindent Consequently, for any labeled Motzkin tree $t$, there is a minimal
natural number $m$ such that $t$ is an $m$-open $\lambda$-term. It can be
computed for instance by the following function.

\coqlisting{minimal_openness}{label.v}

\noindent The function \coq{skeleton} and the relation \coq{label} are pseudo-inverses in
the sense of the following two lemmas.

\coqlisting{label_skeletonK}{label.v}

\coqlisting{skeleton_labelK}{label.v}

The lemmas \coq{label_skeletonK} and \coq{skeleton_is_open_eq} jointly establish
that the labeled Motzkin tree \coq{t} is a \coq{(minimal_openness t)}-open
$\lambda$-term.

\coqlisting{lmt_minimal_openness}{label.v}

\noindent Finally, it is easy to prove by induction that \coq{minimal_openness t}
indeed computes the smallest openness $m$ such that \coq{t} is an $m$-open
$\lambda$-term.

\coqlisting{minimality}{label.v}

\section{Use Cases}
\label{bt17-sec}

In this section we use the previous examples of types to formalize all the
propositions in Bodini and Tarau's work~\cite{lopstr17} that are related to Motzkin trees and pure
$\lambda$-terms.

\vspace{-8pt}
\subsection{Another definition for closable skeletons}
\label{closable2sec}

Bodini and Tarau~\cite[section 3]{lopstr17} first defined closable skeletons
with a Prolog predicate -- named \coq{isClosable} -- whose adaptation in Coq is
\coqlisting{isClosable}{BT17.v}
For each $\lambda$ binder this function increments a counter \coq{V} (starting at
\coq{0}). Then it checks at each leaf that its label is strictly positive. This
definition is slightly more complicated than that of the Coq predicate
\coq{is_closable} presented in Sect.~\ref{sec:closable}. We have proved formally
that both definitions are equivalent:
\coqlisting{is_closable_isClosable_eq}{BT17.v}
\noindent The two implications of this equivalence are proved by structural
induction and thanks to the following two lemmas, themselves proved by
structural induction.

\begin{lstlisting}[style=coq]
Lemma isClosable2_S : forall m n, isClosable2 m n -> isClosable2 m (S n).
Lemma isClosable_l : forall m, isClosable (l m).
\end{lstlisting}

\noindent We can notice that this proof is simpler than expected: Although the
generalization \coq{isClosable2} is required to define the predicate
\coq{isClosable}, the proof avoids the effort to invent generalizations to
\coq{isClosable2} of the predicate \coq{is_closable} and the equivalence lemma.
Similarly, after ``packing'' the predicate \coq{isClosable} in the following
record type, it was possible to define and prove isomorphism with the algebraic
datatype \coq{closable} without having to generalize the record and the
datatype to \coq{isClosable2}.

\coqlisting{recClosable}{BT17.v}

\vspace{-8pt}
\subsection{Two definitions for the size of terms}
\label{sizesSec}

Bodini and Tarau~\cite[Proposition 1]{lopstr17} state the following proposition
to justify that two different size definitions lead to the same sequence of
numbers of closed $\lambda$-terms modulo variable renaming, counted by
increasing size.

\begin{proposition}\label{prop1}
The set of terms of size $n$ for size defined by the respective weights 0, 1 and
2 for variables, abstractions and applications is equal to the set of terms of
size $n+1$ for size defined by weight 1 for variables, abstractions and
applications.
\end{proposition}
This proposition holds not only for all Motzkin trees (without labels), but also
for closable ones, labeled ones, and for $m$-open $\lambda$-terms. Since we
proposed two Coq types for closable Motzkin trees and for $m$-open
$\lambda$-terms, we formalize Proposition~\ref{prop1} by six 
propositions in Coq, all of the form

\begin{lstlisting}[style=coq]
Proposition proposition1X : forall t : X, size111X t = size012X t + 1.
\end{lstlisting}

\noindent with \coq{X} in $\{$\coq{motzkin}, \coq{rec_closable}, \coq{closable},
\coq{lmt}, \coq{rec_open}, \coq{open}$\}$, and with adequate functions
\coq{size111X} and \coq{size012X}, not detailed here, defining both sizes for
each type. More precisely, thanks to the coercion (\coq{P_struct :> T}) in the
record types, the functions \coq{size*rec_P} are not defined, but advantageously
replaced by the functions \coq{size*T}. Here, \coq{*} is either \coq{111} or
\coq{012} and (\coq{T},\coq{P}) is either (\coq{motzkin},\coq{closable}) or
(\coq{lmt},\coq{open}). For record types, the proposition then takes the following form:
\begin{lstlisting}[style=coq]
Proposition proposition1rec_P : forall t : rec_P, size111T t = size012T t + 1.
\end{lstlisting}

\noindent It is a straightforward consequence of the corresponding proposition
on the type \coq{T} (named {\small\texttt{proposi\-tion1T}}, according to our naming
conventions). This mechanism being similar for all record types,
it can easily be mechanized.

The situation is very different with -- potentially -- dependent types (named
\coq{P} in our general framework), if we forbid ourselves to use their
isomorphism with a record type to prove their proposition (named
\coq{proposition1P}, according to our naming conventions). Here, the
propositions for \coq{P} in $\{$\coq{closable},\coq{open}$\}$ are proved by
structural induction and linear arithmetic, because the latter suffices to
inductively define the size functions. However, the general situation may be
arbitrarily more complex, so no general mechanization can be
considered.

\vspace{-8pt}
\subsection{Characterization of closable Motzkin trees}
\label{prop2sec}

Bodini and Tarau~\cite[Proposition 2]{lopstr17} also propose the following characteristic
property for closable Motzkin trees.

\begin{proposition}\label{prop2}
A Motzkin tree is the skeleton of a closed $\lambda$-term if and only if it
exists at least one $\lambda$-binder on each path from the leaf to the
root.
\end{proposition}

 After defining a closed
$\lambda$-term as a $0$-open $\lambda$-term, we can state Proposition~\ref{prop2}
in Coq, as follows.

\begin{lstlisting}[style=coq]
Definition is_closed t := is_open 0 t.
Proposition proposition2 : forall mt : motzkin,
  (exists t : lmt, skeleton t = mt /\ is_closed t) <-> is_closable mt.
\end{lstlisting}

This example shows the advantage of having a formalization by restriction of a
more general type \coq{T} (\coq{motzkin} here) with a predicate \coq{is_P}
(\coq{is_closable} here), compared to the precise types \coq{rec_P} and \coq{P}
(\coq{rec_closable} and \coq{closable} here). Indeed, with these last types,
only one of two implications of Proposition~\ref{prop2} is expressible, for
instance with the following lemmas.

\coqlisting{proposition2rec_closable}{BT17.v}
\coqlisting{proposition2closable}{BT17.v}

These proofs are straightforward.

\vspace{-8pt}
\subsection{Characterization of uniquely closable Motzkin trees}
\label{prop4sec}

Bodini and Tarau propose the following characteristic property for uniquely
closable Motzkin trees~\cite[Proposition 4]{lopstr17}.

\begin{proposition}\label{prop4}
A skeleton is uniquely closable if and only if exactly one lambda binder
is available above each of its leaf nodes.
\end{proposition}

The predicates \coq{is_ucs} and \coq{is_ucs_aux} presented in
Section~\ref{sec:uniquely} correspond to the Prolog predicate
\texttt{uniquelyClosable2} of~\cite[Section 4]{lopstr17} and to the
characteristic property ``exactly one lambda binder is available above each of
its leaf nodes'' of Proposition 4 of~\cite{lopstr17}. Therefore, proving
Proposition~\ref{prop4} consists in showing that this property is equivalent to the
definition ``We call a skeleton \emph{uniquely closable} if it exists exactly
one closed lambda term having it as its skeleton.''~\cite[page 6]{lopstr17},
which gives the following Coq code.

\coqlisting{proposition4}{BT17.v}

\noindent However, this proposition cannot be proved directly, because
\coq{(is_closed t)} is a special case of \coq{(is_open m t)}, which is
parametrized by a natural number \coq{m}, while \coq{(is_ucs mt)} is a special
case of of \coq{(is_ucs_aux mt b)}, which is only parameterized by a Boolean
\coq{b}. The rest of this section addresses this issue by generalizing the
proposition to any natural number \coq{m}, using a characterization
\coq{(ucs1_aux mt m)} parametrized by this integer and put in correspondence
with \coq{(is_ucs_aux mt b)}.

The following predicates \coq{ucs1_aux} and \coq{ucs1} adapt the Prolog
predicate \texttt{uniquelyClosable1} from~\cite{lopstr17} in Coq.

\coqlisting{ucs1}{BT17.v}

We then use the predicate \coq{ucs1_aux} to state a generalization of
proposition~\ref{prop4} to any openness $m$, then the predicate \coq{ucs1} to
state its specialization when $m = 0$, which is a variant of
Proposition~\ref{prop4}.

\coqlisting{proposition4ucs1_aux}{BT17.v}

\coqlisting{proposition4ucs1}{BT17.v}

Independently, we can prove that the two charaterizations of uniquely closable
Motzkin trees are equivalent.

\coqlisting{ucs1_is_ucs_eq}{BT17.v}

As usually when formalizing pen-and-paper proofs, we get more precise statements
and more detailed proofs. For example, we formally proved Proposition 1
in~\cite{lopstr17} as four propositions, corresponding to four distinct data
families.

\section{Conclusions and Perspectives}\label{sec:conclusion}

We have presented a framework to define and formally prove isomorphisms
between Coq datatypes, and to produce random generators for them. After applying
it to several examples related to lambda term families, we have formalized in
Coq a large subset of the computational and logical content of Bodini and
Tarau's paper~\cite{lopstr17} about pure $\lambda$-terms. 

Technically, our present approach using interfaces allows us to
automatically derived only one of two round-trip properties, that
state that the considered transformations are inverse bijections. The
other one, which proceeds by induction on the type \coq{P}, cannot be
generated automatically by a functor, however, we can prove it
automatically using some advanced tactic combinations using Ltac. 
In the near future, we plan to investigate in more details
whether using external tools like
MetaCoq~\cite{DBLP:journals/jar/SozeauABCFKMTW20} or
elpi~\cite{DBLP:conf/lpar/DunchevGCT15} and
Coq-elpi~\cite{tassi:hal-01637063} would increase the genericity of
our approach compared to simply relying on Ltac.

Our framework obviously applies to other formalization topics. It was inspired
by previous work, including one on Coq representations of permutations and
combinatorial maps~\cite{dg18}. We plan to  complete this work and revisit it
using this structuring framework. The proofs of isomorphisms presented in this
paper were elementary because the two types in bijection were very close to one another.
In the more general case of two different points of view on
the same family (e.g., permutations seen as injective endofunctions or products
of disjoint cycles), isomorphisms can be arbitrarily more difficult to prove.

\section*{Acknowledgments}
\label{sec:acknowledgments}

This work has been supported by the EIPHI Graduate School (contract ANR-17-EURE-0002).


\begin{thebibliography}{10}

\bibitem{BBD19}
Maciej Bendkowski, Olivier Bodini, and Sergey Dovgal.
\newblock Statistical {Properties} of {Lambda} {Terms}.
\newblock {\em The Electronic Journal of Combinatorics}, pages P4.1--P4.1,
  October 2019.
\newblock \href {https://doi.org/10.37236/8491} {\path{doi:10.37236/8491}}.

\bibitem{BC04}
Yves Bertot and Pierre Cast\'eran.
\newblock {\em {I}nteractive {T}heorem {P}roving and {P}rogram {D}evelopment,
  \mbox{{C}oq'{A}rt : {T}he} {C}alculus of {I}nductive {C}onstructions}.
\newblock Springer-Verlag, Berlin/Heidelberg, May 2004.
\newblock 469 pages.

\bibitem{lopstr17}
Olivier Bodini and Paul Tarau.
\newblock On uniquely closable and uniquely typable skeletons of lambda terms.
\newblock In Fabio Fioravanti and John~P. Gallagher, editors, {\em Logic-Based
  Program Synthesis and Transformation - 27th International Symposium, {LOPSTR}
  2017, Namur, Belgium, October 10-12, 2017, Revised Selected Papers}, pages
  252--268, 2017.

\bibitem{quick2000}
Koen~Claessen and John~Hughes.
\newblock {QuickCheck}: a lightweight tool for random testing of {Haskell}
  programs.
\newblock In {\em Proceedings of the Fifth ACM SIGPLAN International Conference
  on Functional Programming}, volume~35 of {\em SIGPLAN Not.}, pages 268--279.
  ACM, New York, 2000.
\newblock \href {https://doi.org/10.1145/351240.351266}
  {\path{doi:10.1145/351240.351266}}.

\bibitem{DBLP:journals/flap/CohenCHM17}
Cyril Cohen, Thierry Coquand, Simon Huber, and Anders M{\"{o}}rtberg.
\newblock Cubical type theory: {A} constructive interpretation of the
  univalence axiom.
\newblock {\em {FLAP}}, 4(10):3127--3170, 2017.
\newblock \url{http://collegepublications.co.uk/ifcolog/?00019}.

\bibitem{DBLP:conf/cpp/CohenDM13}
Cyril Cohen, Maxime D{\'{e}}n{\`{e}}s, and Anders M{\"{o}}rtberg.
\newblock Refinements for free!
\newblock In Georges Gonthier and Michael Norrish, editors, {\em Certified
  Programs and Proofs - Third International Conference, {CPP} 2013, Melbourne,
  VIC, Australia, December 11-13, 2013, Proceedings}, volume 8307 of {\em
  LNCS}, pages 147--162. Springer, 2013.
\newblock \href {https://doi.org/10.1007/978-3-319-03545-1\_10}
  {\path{doi:10.1007/978-3-319-03545-1\_10}}.

\bibitem{coqmanual}
{Coq development team}.
\newblock {\em {The Coq Proof Assistant Reference Manual, Version 8.13.2}}.
\newblock INRIA, 2021.
\newblock \url{http://coq.inria.fr}.

\bibitem{de_bruijn_lambda_1972}
N.~G. de~Bruijn.
\newblock Lambda calculus notation with nameless dummies, a tool for automatic
  formula manipulation, with application to the {Church}-{Rosser} theorem.
\newblock {\em Indagationes Mathematicae (Proceedings)}, 75(5):381--392,
  January 1972.
\newblock \href {https://doi.org/10.1016/1385-7258(72)90034-0}
  {\path{doi:10.1016/1385-7258(72)90034-0}}.

\bibitem{dg18}
Catherine Dubois and Alain Giorgetti.
\newblock Tests and proofs for custom data generators.
\newblock {\em Formal Aspects of Computing}, 30:659--684, Jul 2018.
\newblock \href {https://doi.org/10.1007/s00165-018-0459-1}
  {\path{doi:10.1007/s00165-018-0459-1}}.

\bibitem{DBLP:conf/lpar/DunchevGCT15}
Cvetan Dunchev, Ferruccio Guidi, Claudio~Sacerdoti Coen, and Enrico Tassi.
\newblock {ELPI: Fast, Embeddable, {$\lambda$}Prolog Interpreter}.
\newblock In Martin Davis, Ansgar Fehnker, Annabelle McIver, and Andrei
  Voronkov, editors, {\em Logic for Programming, Artificial Intelligence, and
  Reasoning - 20th International Conference, {LPAR-20} 2015, Suva, Fiji,
  November 24-28, 2015, Proceedings}, volume 9450 of {\em LNCS}, pages
  460--468. Springer, 2015.
\newblock \href {https://doi.org/10.1007/978-3-662-48899-7\_32}
  {\path{doi:10.1007/978-3-662-48899-7\_32}}.

\bibitem{GL13}
Katarzyna Grygiel and Pierre Lescanne.
\newblock Counting and generating lambda terms.
\newblock {\em Journal of Functional Programming}, 23(5):594--628, September
  2013.
\newblock \href {https://doi.org/10.1017/S0956796813000178}
  {\path{doi:10.1017/S0956796813000178}}.

\bibitem{Pierce:SF4}
Leonidas Lampropoulos and Benjamin~C. Pierce.
\newblock {\em {QuickChick}: Property-Based Testing in Coq}.
\newblock Software Foundations series, volume 4. Electronic textbook, August
  2022.
\newblock Version 1.3.1. \url{https://softwarefoundations.cis.upenn.edu/qc-1.3.1}.

\bibitem{Les13}
Pierre Lescanne.
\newblock On counting untyped lambda terms.
\newblock {\em Theoretical Computer Science}, 474:80--97, February 2013.
\newblock \href {https://doi.org/10.1016/j.tcs.2012.11.019}
  {\path{doi:10.1016/j.tcs.2012.11.019}}.

\bibitem{DBLP:conf/tphol/Magaud03}
Nicolas Magaud.
\newblock Changing data representation within the Coq system.
\newblock In David~A. Basin and Burkhart Wolff, editors, {\em Theorem Proving
  in Higher Order Logics, 16th International Conference, TPHOLs 2003, Rome,
  Italy, September 8-12, 2003, Proceedings}, volume 2758 of {\em LNCS}, pages
  87--102. Springer, 2003.
\newblock \href {https://doi.org/10.1007/10930755\_6}
  {\path{doi:10.1007/10930755\_6}}.

\bibitem{DBLP:conf/types/McBride00}
Conor McBride.
\newblock Elimination with a motive.
\newblock In Paul Callaghan, Zhaohui Luo, James McKinna, and Robert Pollack,
  editors, {\em Types for Proofs and Programs, International Workshop, {TYPES}
  2000, Durham, UK, December 8-12, 2000, Selected Papers}, volume 2277 of {\em
  LNCS}, pages 197--216. Springer, 2000.
\newblock \href {https://doi.org/10.1007/3-540-45842-5\_13}
  {\path{doi:10.1007/3-540-45842-5\_13}}.

\bibitem{DBLP:journals/jfp/McBrideM04}
Conor McBride and James McKinna.
\newblock The view from the left.
\newblock {\em J. Funct. Program.}, 14(1):69--111, 2004.
\newblock \href {https://doi.org/10.1017/S0956796803004829}
  {\path{doi:10.1017/S0956796803004829}}.

\bibitem{10.1145/3519939.3523707}
Zoe Paraskevopoulou, Aaron Eline, and Leonidas Lampropoulos.
\newblock Computing correctly with inductive relations.
\newblock In {\em Proceedings of the 43rd ACM SIGPLAN International Conference
  on Programming Language Design and Implementation}, PLDI 2022, page
  966–980, New York, NY, USA, 2022. Association for Computing Machinery.
\newblock \href {https://doi.org/10.1145/3519939.3523707}
  {\path{doi:10.1145/3519939.3523707}}.

\bibitem{DBLP:journals/jar/SozeauABCFKMTW20}
Matthieu Sozeau, Abhishek Anand, Simon Boulier, Cyril Cohen, Yannick Forster,
  Fabian Kunze, Gregory Malecha, Nicolas Tabareau, and Th{\'{e}}o Winterhalter.
\newblock The \textsc{MetaCoq} project.
\newblock {\em J. Autom. Reason.}, 64(5):947--999, 2020.
\newblock \href {https://doi.org/10.1007/s10817-019-09540-0}
  {\path{doi:10.1007/s10817-019-09540-0}}.

\bibitem{tassi:hal-01637063}
Enrico Tassi.
\newblock {Elpi: an extension language for Coq (Metaprogramming Coq in the Elpi
  {$\lambda$}Prolog dialect)}.
\newblock The Fourth International Workshop on Coq for Programming Languages:
  CoqPL 2018, January 2018.
\newblock \url{https://hal.inria.fr/hal-01637063}.

\bibitem{hottbook}
The {Univalent Foundations Program}.
\newblock {\em Homotopy Type Theory: Univalent Foundations of Mathematics}.
\newblock \url{https://homotopytypetheory.org/book}, Institute for Advanced
  Study, 2013.

\bibitem{DBLP:conf/popl/Wadler87}
Philip Wadler.
\newblock Views: {A} way for pattern matching to cohabit with data abstraction.
\newblock In {\em 14th {ACM} Symposium on Principles of Programming Languages,
  Munich, Germany, January 21-23, 1987}, pages 307--313. {ACM} Press, 1987.
\newblock \href {https://doi.org/10.1145/41625.41653}
  {\path{doi:10.1145/41625.41653}}.

\end{thebibliography}
\end{document}